# Multi-ligand simultaneous docking analysis of Moringa oleifera phytochemicals reveals enhanced BCL-2 inhibition via synergistic action

Asmita Saha
*Department of Biological Sciences*
*Dayananda Sagar University*
Bengaluru, India
0009-0002-1416-0665

Belaguppa Manjunath Ashwin Desai
*Department of ECE*
*Dayananda Sagar University*
Bengaluru, India
0000-0002-4144-6266

Pronama Biswas
*Department of Biological Sciences*
*Dayananda Sagar University*
Bengaluru, India
0000-0003-2950-1530

*Abstract*— *Moringa oleifera*, known for its medicinal properties, contains bioactive compounds such as polyphenols and flavonoids with diverse therapeutic potentials, including anti-cancer effects. This study investigates the efficacy of M. oleifera leaf phytochemicals in inhibiting BCL-2, a critical protein involved in cancer cell survival. For the first time, multi-ligand simultaneous docking (MLSD) has been employed to understand the anti-cancer properties of M. oleifera leaf extract. Molecular docking techniques, including single-ligand and MLSD, were used to assess binding interactions with BCL-2. Single-ligand docking revealed strong binding affinities for compounds such as niazinin, alpha carotene, hesperetin, apigenin, niaziminin B, and niazimicin A, with some compounds even surpassing Venetoclax, a commercial BCL-2 inhibitor. MLSD highlighted inter-ligand interactions among apigenin, hesperetin, and niazimicin A, exhibiting a binding affinity of -14.96 kcal/mol, indicating a synergistic effect. These results shed light on the potential synergistic effects of phytochemicals when using multi-ligand simultaneous docking, underscoring the importance of considering compound interactions in the development of therapeutic strategies.

## I. Introduction

*Moringa oleifera*, commonly known as drumstick, is a member of the Moringaceae family and has been valued for its anti-cancer, anti-diabetic, anti-microbial, and anti-inflammatory properties [1]. For instance, *M. oleifera* leaf extracts have shown a significant reduction in the viability of MCF-7 breast cancer cells in MTT assays [2]. Computational simulations have also identified specific compounds within the extracts as potential CDK-2 inhibitors, further supporting their therapeutic potential [2]. Studies on A549 cells revealed that *M. oleifera* leaf extracts significantly decrease BCL-2 expression in a dose-dependent manner, as demonstrated by real-time PCR, indicating their potential to induce apoptosis [3]. Additionally, four major compounds uniquely found in *M. oleifera*—niazimicin, niaziminin, niazinin, and niazirin exhibit promising anticancer properties and have been investigated for their potential to lower blood pressure [4].

The BCL-2 family of proteins, including anti-apoptotic members (BCL-2, BCL-XL) and pro-apoptotic members (BAX, BAK), plays a critical role in regulating the intrinsic apoptotic pathway. Anti-apoptotic proteins inhibit apoptotic signals by binding to the mitochondrial membrane, thus promoting cell survival [5]. Elevated levels of BCL-2 promote resistance to apoptosis and unchecked cell proliferation, making it an important prognostic marker in ER-positive tumors and is overexpressed in triple-negative breast cancer (TNBC) and luminal A subtype breast cancer [6]. Several studies have demonstrated that targeting BCL-2 is an effective therapeutic strategy in cancer, with Venetoclax-based therapies being widely used, alongside ongoing investigations into phytochemicals that can also inhibit BCL-2 [7]. Venetoclax, an FDA-approved drug, effectively targets BCL-2 in cancers such as chronic lymphocytic leukemia (CLL), small lymphocytic lymphoma (SLL), and acute myeloid leukemia (AML). It binds to the BH3 domain of BCL-2, inhibiting its function and inducing apoptosis by releasing regulatory proteins that activate pro-apoptotic factors like BAX. However, mutations in BCL-2, such as Gly101Val and Asp103Tyr, can lead to resistance against Venetoclax and contribute to cancer recurrence [8]. This advocates the need for finding new inhibitors for BCL-2.

To understand how bioactive compounds in *M. oleifera* may inhibit BCL-2, it is essential to study their interactions with this target protein. Molecular docking, an in-silico technique, predicts the binding affinity of drugs to proteins by evaluating various ligand conformations and identifying the optimal binding mode based on energy scores. In real-world scenarios, binding processes often include multiple participants such as substrates, cofactors and ions, all of which significantly influence the interactions. Traditional docking methods, which dock one ligand at a time, fail to capture these intricate interactions, resulting in compromised docking modes and affinity predictions. Multi-ligand simultaneous docking (MLSD) addresses this gap by enabling the simultaneous docking of multiple ligands, thus mimicking the nearly true molecular recognition processes. This approach not only enhances our understanding of competitive and cooperative binding scenarios but also reveals inter-ligand interactions that can imply synergistic effects. Consequently, MLSD offers a more detailed view of how multiple bioactive compounds may work together to inhibit a target protein effectively [9]. Raghavendra et al. were among the first to employ Multiple Ligand Simultaneous Docking (MLSD) to demonstrate multiple substrate binding on the Polyphenoloxidase protein, highlighting that MLSD can better mimic real molecular interactions compared to traditional docking methods [10]. Studies conducted by Gupta et al. provided insights into developing potential drug combinations against SARS-CoV-2 protease using MLSD. They predicted potential synergistic effects and stable interactions between the approved drug 2-deoxy-D-glucose and the phytochemical Ruxolitinib, leading to enhanced inhibition of the virus [11]. Similarly, Devi et al. applied MLSD to study the combined effects of Epigallocatechin-O-Gallate and Withaferin A on skin-aging enzymes, establishing the potential for combinatorial therapy [12]. Although MLSD is still a relatively new approach and fewer docking studies have used this technique, these studies emphasize its superiority in capturing complex molecular interactions.

In this study, we employed both single-ligand molecular docking and MLSD to investigate the potential synergistic effects of phytochemicals present in *M. oleifera* leaf extracts. Through MLSD, we gain deeper insights into the collective binding behaviors and conformational changes induced in multiple ligands, offering a more comprehensive perspective on their possible synergistic inhibition of BCL-2.

## II. MATERIALS AND METHODS

### A. Protein Acquisition and validation

The three-dimensional structure of BCL-2 was retrieved from RCSB PDB (http://www.rcsb.org/pdb/) with the PDB ID 4IEH. The protein quality was assessed using "PROCHECK" (https://saves.mbi.ucla.edu/) and "VoroMQA" (https://bioinformatics.lt/wtsam/voromqa). In PROCHECK, we used "Whatcheck", "Verify 3D" and "Procheck" tools. The structure underwent rigorous assessment, ensuring a Ramachandran score of over 90% [13], "Verify 3D" value exceeding 80%, and "Voromqa" Global plot position above the lowest 5% threshold (indicated by the red line) [14].

### B. Ligand Selection and Acquisition

A total of 31 phytochemicals were identified from *M. oleifera* leaf extracts using the Indian Medicinal Plants Phytochemistry and Therapeutics (IMPPAT) database (https://cb.imsc.res.in/imppat/home). Each phytochemical underwent Pan-Assay Interference Compounds (PAINS) analysis using ADMETlab 2.0 (https://admetmesh.scbdd.com/) [15]. This validation step was essential to avoid false positives and reduce the likelihood of multiple target binding, as our primary goal is to identify specific BCL-2 inhibitors with minimal non-specific interactions [12]. The compounds vanillin, apigenin, trigonelline, ascorbic acid, hesperetin, phenylacetonitrile, palmitic acid, niazinin, niazimicin A, niaziminin B, niazirin, alpha carotene, ferulic acid, 4-hydroxycinnamic acid, 1,7-Octadien-3-ol, 2,6-dimethyl-, and N,alpha-L-rhamnopyranosyl vincosamide did not show PAINS alerts. Therefore, these compounds were selected for docking analysis to ensure that we focused on specific interactions with BCL-2. Their three-dimensional structures were retrieved from PubChem (http://pubchem.ncbi.nlm.nih.gov) in Structure Data File (SDF) format, with Compound IDs, 1183, 5280443, 5570, 54670067, 72281, 8794, 985, 10088810, 10247749, 44559760, 129556, 6419725, 445858, 637542, 90789, and 71717770, respectively. Additionally, the structure of the commercial BCL-2 inhibitor used for validation was obtained from PubChem in SDF format with Compound ID 49846579 (Venetoclax).

### C. Tools and Software

Open Babel (http://openbabel.org) was utilized on Ubuntu to convert the ligand SDF files to Protein Data Bank, Partial Charge, and Atom Type (PDBQT) file format for docking. The protein was prepared using AutoDockTools-1.5.7 (https://autodock.scripps.edu/), and the output was saved as PDBQT files. Molecular docking was performed using the Vina Script method within a Conda environment accessed through Visual Studio Code. The resulting PDBQT files were subsequently analyzed in Discovery Studio 2021 Client (https://www.3ds.com/products/biovia/discovery-studio/visualization) and UCSF Chimera 1.17 (https://www.cgl.ucsf.edu/chimera/) to examine amino acid interactions and binding pockets. Docking studies were conducted on a server featuring an AMD EPYC 7742 64-Core Processor, 503 GiB of RAM, and NVIDIA A100 GPUs, running Ubuntu 22.04.4 LTS with a 6.5.0-41-generic kernel and 1.8 TB primary disk. The server was accessed via an HP Pavilion laptop equipped with a 12th Gen Intel Core i5 CPU and Intel Iris Xe Graphics GPU.

### D. Molecular Docking

BCL-2 was prepared using AutoDockTools which involved removal of co-crystallised ligands and water molecules, addition of hydrogen and gasteiger charges and repairing missing atoms. The resulting receptor PDB file was saved in PDBQT format as "receptor.pdbqt". Blind docking was performed by setting the grid box to accommodate the entire receptor [16]. The grid dimensions were calculated using UCSF Chimera and saved as "config.txt". For ligand preparation, the obabel command was used in Ubuntu to add hydrogens and gasteiger charges. Energy minimization was further performed on the ligand structure and converted to PDBQT files.

The selected ligands, including the commercial inhibitor, were individually docked with BCL-2 using the Vina scoring function in Visual Studio Code. A dedicated Conda environment was set up in VS Code to facilitate the docking sessions via batch processing commands. Biovia and chimera were used to analyse the docking scores, binding pockets and compared to Venetoclax. The standard deviation of the binding affinities of single docking and MLSD varied in the range of ±0.3–0.5. In MLSD, two ligands were docked simultaneously to the receptor, allowing the ligands to sense each other and interact with the receptor accordingly. For visualization in Biovia Visualizer, the ligands were individually selected, copied, and pasted into the 'receptor.pdbqt' tab. The 2D amino acid interactions of each ligand was checked separately. A similar approach was used in UCSF Chimera to better visualize the binding 3D binding pocket. The schematic representation of the steps followed has been illustrated in Fig. 1.

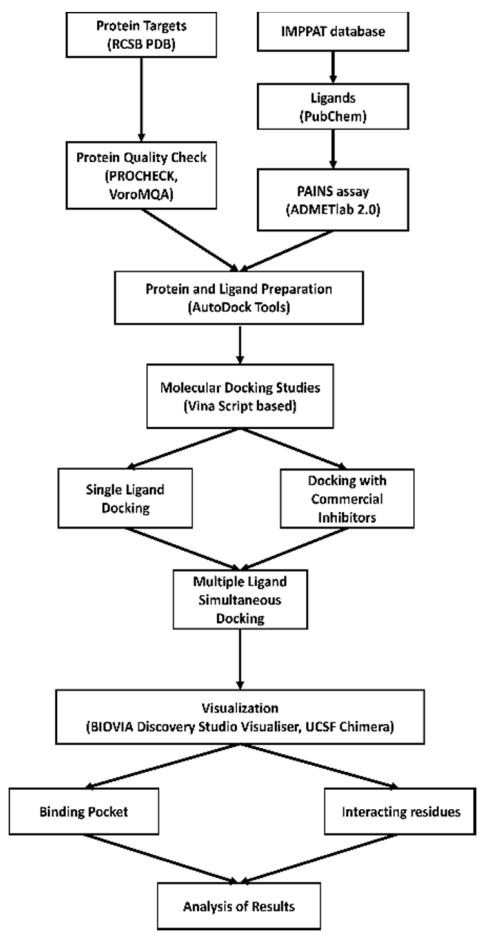

Fig. 1. Methodology followed for docking

## III. RESULTS

### A. Single ligand docking revealed potential synergistic phytochemicals

The strongest binding affinity was observed with niazinin at -10.19 kcal/mol, indicating a very strong docking interaction, as more negative binding affinities signify stronger binding. The 3D pocket visualization revealed that niazinin binds in a different pocket than Venetoclax. Alpha carotene followed with a binding affinity of -9.54 kcal/mol, binding perfectly in the same pocket as Venetoclax. However, due to its large size, it posed challenges for synergistic effects with other molecules leading to its exclusion from MLSD considerations. Other compounds such as N,alpha-L-rhamnopyranosyl vincosamide (-8.94 kcal/mol), hesperetin (-7.24 kcal/mol), apigenin (-6.94 kcal/mol), niaziminin B (-6.91 kcal/mol) and niazimicin A (-6.72 kcal/mol) exhibited good binding affinities and were observed to bind near Venetoclax's binding pocket. The results of single ligand docking are shown in Fig. 2.

For MLSD, ligand combinations were strategically selected based on their binding positions (Table I). Combinations such as C1 and C2 were chosen because each ligand in the combination occupied distinct parts of the Venetoclax binding pocket, as determined from single docking results. The rationale behind selecting these combinations was that by covering different parts of the binding site, the ligands could potentially bind simultaneously, allowing for inter-ligand interactions that might lead to synergistic effects and enhanced stability within the pocket. Another selected combination was C3, as both ligands docked in nearly the same binding pocket with similar binding affinities.

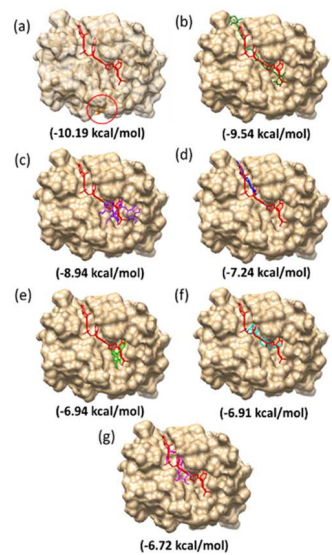

Fig. 2. Single ligand docking binding pockets of *Moringa oleifera* phytochemicals compared to Venetoclax (red). The structures of *M. oleifera* phytochemicals are superimposed with the docked structure of Venetoclax for comparison. (a) Niazinin (orange) binds in a different pocket than Venetoclax. The binding pocket is highlighted with a red circle, and the BCL-2 structure is made transparent for easier visualization. (b) Alpha carotene (dark green) binds in the same pocket as Venetoclax. The binding sites of other phytochemicals, including (c) N,alpha-L-rhamnopyranosyl vincosamide, (d) Hesperetin, (e) Apigenin, (f) Niaziminin B, and (g) Niazimicin A, are located in different parts of the Venetoclax binding site. These phytochemicals were selected for further MLSD due to their potential for synergistic interactions.

Further MLSD was conducted using three-ligand combinations of C4, C5, C6, and C7. Niazimicin A and niaziminin B were included in these combinations as they could be used with potential linkers, aiming to explore synergistic effects and optimal binding configurations. This approach of selecting combinations may be limited due to lack of sampling and absence of molecular dynamic simulations, which could result in some potential combinations being overlooked. This may introduce some degree of bias.

### B. MLSD reveals potential synergistic effects

MLSD revealed that few of the chosen combinations exhibited enhanced binding affinities compared to single ligand docking and docking with commercial inhibitor, Venetoclax. The three-ligand combination of C4 showed significant inter-ligand interactions (Fig. 3). These ligands fit perfectly in the Venetoclax binding pocket, forming hydrogen bonds, pi-pi, and pi-alkyl interactions between hesperetin and apigenin, and a hydrogen bond between apigenin and niazimicin A. The binding affinity of this combination is -

14.96 kcal/mol, which is higher than Venetoclax's -11.44 kcal/mol. TYR161, ARG66, ASP62, GLY104, and GLN58, which form hydrogen bonds with Venetoclax, also interact with the combination of ligands, forming pi bonds, electrostatic interactions, and additional hydrogen bonds. This further validates their similar binding pocket to the commercial inhibitor.

TABLE I. COMBINATIONS CHOSEN FOR MLSD ANALYSIS

| Combination | Phytochemicals | Binding Affinity (Kcal/mol) |
|---|---|---|
| C1 | Apigenin + Hesperetin | -12.75 |
| C2 | Hesperetin + N,alpha-L-rhamnopyranosyl vincosamide | -13.05 |
| C3 | Niazimicin A + Niaziminin B | -9.95 |
| C4 | Apigenin + Hesperetin + Niazimicin A | -14.96 |
| C5 | Apigenin + Hesperetin + Niaziminin B | -14.86 |
| C6 | Hesperetin + N, alpha-L-rhamnopyranosyl vincosamide + Niazimicin A | -12.81 |
| C7 | Hesperetin + N,alpha-L-rhamnopyranosyl vincosamide + Niaziminin B | -13.51 |

In the C5 combination, apigenin and hesperetin bind in the Venetoclax pocket but do not cover the entire pocket, while niaziminin B binds to a completely different pocket. The overall binding affinity of this combination is -14.86 kcal/mol, also higher than the commercial inhibitor. GLY104 and TYR161 amino acids form hydrogen and pi bonds with apigenin and hesperetin respectively (Fig. 4).

Another promising result was observed with the C6 combination, where inter-ligand interactions were seen between N,alpha-L-rhamnopyranosyl vincosamide and niazimicin a. The binding affinity of this combination is -12.81 kcal/mol, surpassing the affinity of the commercial inhibitor. In this case, TYR161 forms hydrogen bonds with both niazimicin A and N, alpha-L-rhamnopyranosyl vincosamide, while GLY104 forms a carbon-hydrogen bond with N,alpha-L-rhamnopyranosyl vincosamide. These interactions are consistent with those observed in Venetoclax. (Fig. 5).

Synergism occurs when the combined effect of multiple compounds surpasses their individual effects, often through the formation of stable intermolecular complexes or the creation of dimers or new phenolic products [17]. This is evident from our docking results, where the multi-ligand docked complex shows a stronger binding affinity compared to the single dockings of each phytochemical. Additionally, the interactions within our multi-ligand complexes reveal a stable conformation at the target binding site. These observations suggest a synergistic binding effect of our ligands.

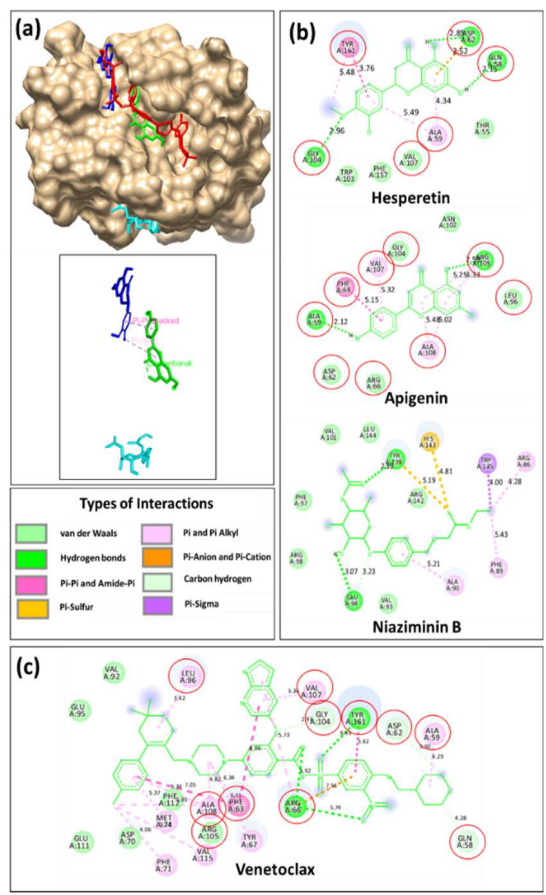

Fig. 3. MLSD results for C4 combination with Hesperetin (blue), Apigenin (green), and Niazimicin A (purple). (a) The docked structures of these ligands are superimposed with Venetoclax (red), showing that the ligand combination coincides with the Venetoclax binding pocket. The inter-ligand interactions between the three M. oleifera phytochemicals have been shown in the box separately for better visualisation. (b) Amino acid interactions between BCL-2 and the phytochemicals. (c) Amino acid interactions between BCL-2 and Venetoclax. The red circles highlight the shared amino acids between Venetoclax and the combination ligands, confirming that they bind to similar pockets.

IV. DISCUSSION

Our study revealed significant synergistic interactions among several compounds, underscoring the potential of *M. oleifera* extract as an effective anti-cancer agent. Specifically, the phytochemicals apigenin, hesperetin, niazimicin A, and N, alpha-L-rhamnopyranosyl vincosamide were identified as key contributors to the inhibitory effect on BCL-2 when used in various combinations. Single-ligand docking results indicated that niazinin had the highest binding affinity (-10.19 kcal/mol) among the tested compounds. However, it bound in a different pocket than Venetoclax and was therefore not considered for further MLSD. MLSD results were particularly compelling, with combinations such as C4 showing a binding affinity of -14.96 kcal/mol, significantly surpassing Venetoclax (-11.44 kcal/mol). This combination exhibited strong inter-ligand interactions, including hydrogen bonds and pi-alkyl interactions with BCL-2. Hydrogen bonds are one of the strongest non-covalent interactions which form between the hydroxyl groups of the ligand and the amino acids of the protein. An increase in hydrogen bonds is therefore a key determinant of binding efficiency and stability. Pi bonds, on the other hand, facilitate charge transfer, which is essential for embedding the ligand into the protein's binding pocket. Another effective combination included C6, with a binding affinity of -12.81 kcal/mol.

enhances the inhibitory potential of the ligand combinations, making them more effective in targeting BCL-2.

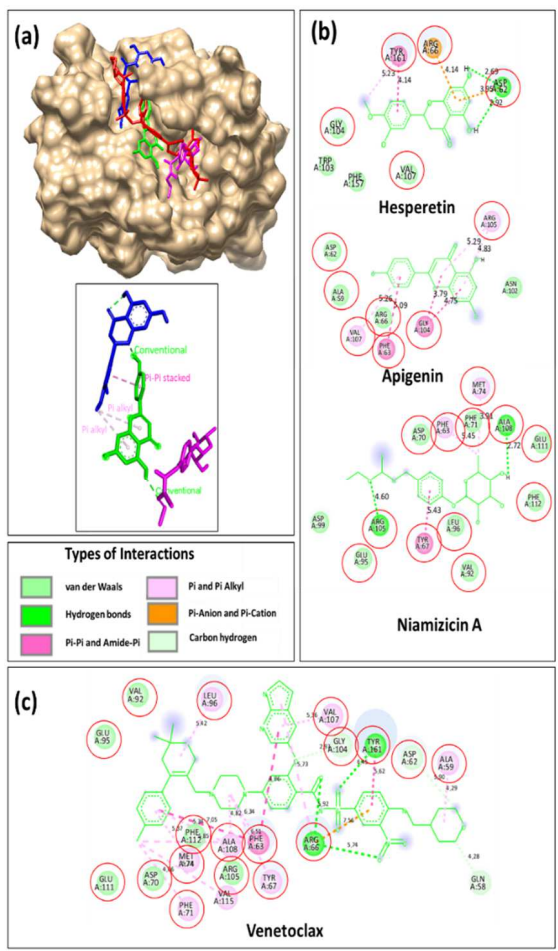

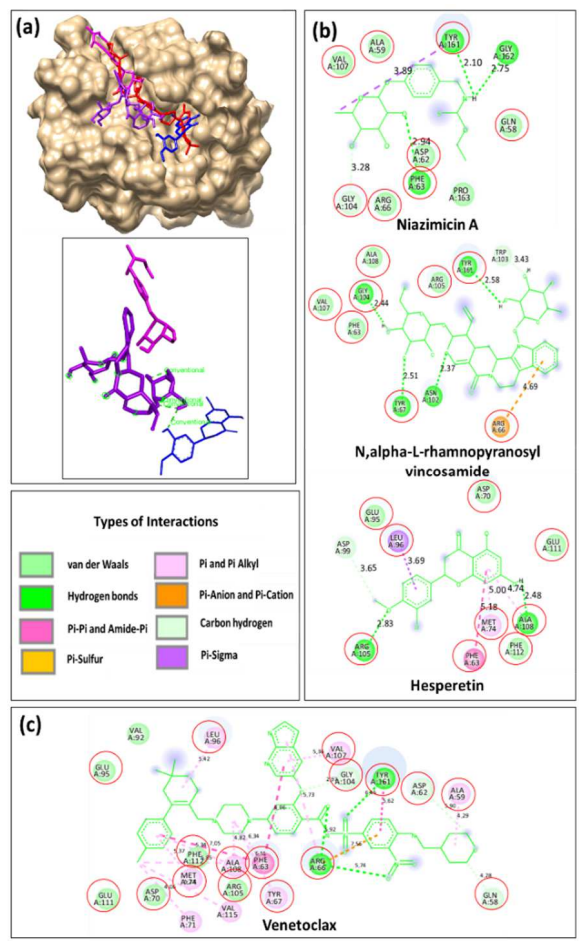

Fig. 4. MLSD results for Hesperetin (blue), Apigenin (green), and Niaziminin B (cyan) in combination. (a) The docked structures of these ligands are superimposed with Venetoclax (red), showing that two ligands coincide with the Venetoclax binding pocket while the second ligand binds to a different pocket. The inter-ligand interactions between the three *M. oleifera* phytochemicals have been shown in the box for better visualisation. (b) Amino acid interactions between BCL-2 and the phytochemicals. (c) Amino acid interactions between BCL-2 and Venetoclax. The red circles highlight the shared amino acids between Venetoclax and the combination ligands, confirming that they bind to similar pockets.

The enhanced binding affinities observed in the MLSD studies are attributed to inter-ligand interactions, including hydrogen bonds, pi-pi, and pi-alkyl interactions. These interactions can stabilize the ligand complexes within the BCL-2 binding pocket, reducing the likelihood of ligand detachment over time and preventing substrate binding [18]. Molecular dynamic simulations, used to analyze the trajectory and root mean square deviation of the complex in a solvated system, can confirm this stability. This increased stability

Fig. 5. MLSD results for Niazimicin A (pink), N,alpha-L-rhamnopyranosyl vincosamide (purple), and Hesperetin (blue) in combination. (a) The docked structures of these ligands are superimposed with Venetoclax (red), showing that all ligands in combination coincide with the Venetoclax binding pocket while the second ligand binds to a different pocket. The hydrogen bond interactions are seen only between two ligands. (b) Amino acid interactions between BCL-2 and the phytochemicals. (c) Amino acid interactions between BCL-2 and Venetoclax. The common amino acids between Venetoclax and the combination ligands, indicated by red circles, validate their similar binding pockets.

These findings suggest that the phytochemicals from *M. oleifera* leaves, particularly when used in combination, have a strong potential to inhibit BCL-2, a key protein involved in cancer cell survival. The observed synergistic effects highlight the promise of these natural compounds in enhancing the efficacy of cancer treatments. The use of natural compounds such as those found in *M. oleifera* offers several advantages over synthetic drugs, including lower toxicity and fewer side effects. The ability of these phytochemicals to inhibit BCL-2 suggests potential applications in developing new, more effective cancer therapies, particularly for cancers resistant to existing treatments like Venetoclax.

A similar study conducted by Li et al. using natural compounds from a Chinese herbal formula to inhibit M$^{pro}$, the main protease of SARS-CoV-2 [18]. They used multi-ligand sequential docking to establish synergistic effects through inter-ligand interactions. Sequential docking, however, has limitations when multiple drugs are administered simultaneously, as it does not account for the order in which drugs bind to a protein. In contrast, MLSD accounts for ligand binding based not only on their affinities but also on their interactions, conformational orientation, and the shape of the protein pocket, dynamically adjusting to the presence of other ligands. This approach allows for a more realistic simulation of how multiple ligands interact with a protein simultaneously, considering various factors that influence binding beyond just affinity.

This study has a few limitations. This study relies on computational models that, while powerful, are inherently predictive and necessitate experimental validation. The assumptions made during docking simulations, such as treating the protein and ligands as static entities, may influence the results. Furthermore, the structures available in databases are limited and might not accurately reflect true conformations. The differences between the computational docking environment and the complexity of biological systems can also impact the predicted effectiveness of drug combinations. The possibility of false positives remains a significant limitation, as these models may predict interactions that do not occur in reality. This Future work should include experimental assays to confirm these findings and address any discrepancies that may arise.

## V. Conclusion

The phytochemicals present in *M. oleifera* leaf extract hold promising potential for cancer therapy by effectively targeting BCL-2, offering a strategy to combat Venetoclax resistance. Apigenin, hesperetin, niazimicin A, and N,alpha-L-rhamnopyranosyl vincosamide exhibited significant binding affinities to BCL-2, surpassing that of the commercial inhibitor Venetoclax. The synergistic interactions observed in MLSD highlight the enhanced inhibitory potential of these compounds when used in combination.

By demonstrating that multiple phytochemicals can work together to inhibit BCL-2 more effectively than a single commercial inhibitor, this research suggests a new avenue for designing multi-targeted therapeutic strategies and fragment-based drug design. Such combination therapies could potentially overcome resistance mechanisms, reduce drug dosages, and minimize side effects, leading to more effective and personalized cancer treatments. However, limitations such as the rigidity of protein receptors in docking analyses can hinder the understanding of conformational changes upon ligand binding.

## Acknowledgment

This study was supported by the Dayananda Sagar University Seed Grant (Grant No. DSU/REG/2022-23/740).

## References

[1] X. Su et al., "*Moringa oleifera* Lam.: a comprehensive review on active components, health benefits and application," *RSC Adv*, vol. 13, no. 35, pp. 24353–24384, 2023, doi: 10.1039/D3RA03584K.

[2] R. Sultan, A. Ahmed, L. Wei, H. Saeed, M. Islam, and M. Ishaq, "The anticancer potential of chemical constituents of Moringa oleifera targeting CDK-2 inhibition in estrogen receptor positive breast cancer using in-silico and in vitro approches," *BMC Complement Med Ther*, vol. 23, no. 1, p. 396, Nov. 2023, doi: 10.1186/s12906-023-04198-z.

[3] K. Bhadresha, V. Thakore, J. Brahmbhatt, V. Upadhyay, N. Jain, and R. Rawal, "Anticancer effect of Moringa oleifera leaves extract against lung cancer cell line via induction of apoptosis," *Advances in Cancer Biology - Metastasis*, vol. 6, p. 100072, Dec. 2022, doi: 10.1016/j.adcanc.2022.100072.

[4] M. Ozturk and K. R. Hakeem, Eds., *Plant and Human Health, Volume 2*. Cham: Springer International Publishing, 2019. doi: 10.1007/978-3-030-03344-6.

[5] D. Kaloni, S. T. Diepstraten, A. Strasser, and G. L. Kelly, "BCL-2 protein family: attractive targets for cancer therapy," *Apoptosis*, vol. 28, no. 1–2, pp. 20–38, Feb. 2023, doi: 10.1007/s10495-022-01780-7.

[6] Y. H. Eom, H. S. Kim, A. Lee, B. J. Song, and B. J. Chae, "BCL2 as a Subtype-Specific Prognostic Marker for Breast Cancer," *J Breast Cancer*, vol. 19, no. 3, p. 252, 2016, doi: 10.4048/jbc.2016.19.3.252.

[7] N. Rahman et al., "Bcl-2 Modulation in p53 Signaling Pathway by Flavonoids: A Potential Strategy towards the Treatment of Cancer," Int J Mol Sci, vol. 22, no. 21, p. 11315, Oct. 2021, doi: 10.3390/ijms222111315.

[8] Y. Xu and H. Ye, "Progress in understanding the mechanisms of resistance to BCL-2 inhibitors," *Exp Hematol Oncol*, vol. 11, no. 1, p. 31, Dec. 2022, doi: 10.1186/s40164-022-00283-0.

[9] H. Li and C. Li, "Multiple ligand simultaneous docking: Orchestrated dancing of ligands in binding sites of protein," *J Comput Chem*, vol. 31, no. 10, pp. 2014–2022, Jul. 2010, doi: 10.1002/jcc.21486.

[10] S. Raghavendra, S. J. Aditya Rao, V. Kumar, and C. K. Ramesh, "Multiple ligand simultaneous docking (MLSD): A novel approach to study the effect of inhibitors on substrate binding to PPO," Comput Biol Chem, vol. 59, pp. 81–86, Dec. 2015, doi: 10.1016/j.compbiolchem.2015.09.008.

[11] A. Gupta, S. S. Chauhan, A. S. Gaur, and R. Parthasarathi, "Computational screening for investigating the synergistic regulatory potential of drugs and phytochemicals in combination with 2-deoxy-D-glucose against SARS-CoV-2," Struct Chem, vol. 33, no. 6, pp. 2179–2193, Dec. 2022, doi: 10.1007/s11224-022-02049-0.

[12] A. Devi et al., "Multiple Ligand Simultaneous Docking Analysis of Epigallocatechin-O-Gallate (Green Tea) and Withaferin A (Ashwagandha) Effects on Skin-Aging Related Enzymes," Indian J Pharm Sci, vol. 85, no. 4, 2023, doi: 10.36468/pharmaceutical-sciences.1171.

[13] R. A. Laskowski, M. W. MacArthur, D. S. Moss, and J. M. Thornton, "PROCHECK: a program to check the stereochemical quality of protein structures," *J Appl Crystallogr*, vol. 26, no. 2, pp. 283–291, Apr. 1993, doi: 10.1107/S0021889892009944.

[14] K. Olechnovič and Č. Venclovas, "VoroMQA web server for assessing three-dimensional structures of proteins and protein complexes," *Nucleic Acids Res*, vol. 47, no. W1, pp. W437–W442, Jul. 2019, doi: 10.1093/nar/gkz367.

[15] S. N. Bolz, M. F. Adasme, and M. Schroeder, "Toward an Understanding of Pan-Assay Interference Compounds and Promiscuity: A Structural Perspective on Binding Modes," *J Chem Inf Model*, vol. 61, no. 5, pp. 2248–2262, May 2021, doi: 10.1021/acs.jcim.0c01227.

[16] S. Forli, R. Huey, M. E. Pique, M. F. Sanner, D. S. Goodsell, and A. J. Olson, "Computational protein–ligand docking and virtual drug screening with the AutoDock suite," Nat Protoc, vol. 11, no. 5, pp. 905–919, May 2016, doi: 10.1038/nprot.2016.051.

[17] M. Olszowy-Tomczyk, "Synergistic, antagonistic and additive antioxidant effects in the binary mixtures," *Phytochemistry Reviews*, vol. 19, no. 1, pp. 63–103, Feb. 2020, doi: 10.1007/s11101-019-09658-4.

[18] H. Li et al., "Multi-ligand molecular docking, simulation, free energy calculations and wavelet analysis of the synergistic effects between natural compounds baicalein and cubebin for the inhibition of the main protease of SARS-CoV-2," *J Mol Liq*, vol. 374, p. 121253, Mar. 2023, doi: 10.1016/j.molliq.2023.121253.